\documentclass[oneside,11pt]{amsart}
\usepackage[english]{babel}
\usepackage{bbm,euscript,amssymb,amsthm,amsmath}
\usepackage{mathrsfs}
\usepackage{amsfonts}
\usepackage{amsmath}
\usepackage{amssymb}
\usepackage{fancybox}
\usepackage{graphicx}
\usepackage{color}
\usepackage{lineno}
\usepackage{pstricks,pstricks-add,pst-math,pst-xkey}
\usepackage{rotating}
\usepackage{amsmath,amsfonts,amssymb,amsthm,graphicx,graphics,epsf}
\usepackage{mathcomp}
\usepackage{textcomp}

\newtheorem{thm}{Theorem} 
\newtheorem{thm*}{Theorem*}
\newtheorem{lemma}[thm]{Lemma} 
\newtheorem{theorem}[thm]{Theorem}

\newtheorem{proposition}[thm]{Proposition}

\def\att{{\mathcal T}}
\def\LL{{\mathcal L}}

\def\ttn{{\log{n}/(\log{\log{n})}}}

\def\ofh{{\overline{fh}}}
\def\oab{{\overline{ab}}}
\def\ojk{{\overline{jk}}}

\def\oad{{\overline{ad}}}
\def\obd{{\overline{bd}}}

\def\ode{{\overline{de}}}
\def\ofj{{\overline{fj}}}

\def\ctwo{{1000}}
\def\cthree{{2040}}

\def\tn{{{\log^{1-\varepsilon} n}}}
\def\tn{{\frac{\log n}{2\log{\log n}}}}

\def\area#1{{\hbox{\rm a}(#1)}}
\def\peri#1{{\hbox{\rm per}(#1)}}
\def\Pr{{\mathbbm P}}

\def\floor#1{{\lfloor{#1}\rfloor}}

\def\EE{{\mathbbm E}}

\def\ignore#1{}
\def\coha#1{{\hbox{\sc Hol}(#1)}}
\def\hol#1{{\coha{#1}}}
\def\real{{\mathbb{R}}}

\title[Large convex holes in random point sets]{Large convex holes in random point
  sets}

\author{J\'ozsef Balogh}
\address{University of Illinois at Urbana-Champaign, USA.}
\email{jobal@math.uiuc.edu}
\thanks{The first author was supported by NSF CAREER Grant DMS-0745185.}
\author{Hern\'an Gonz\'alez-Aguilar}
\address{Facultad de Ciencias, UASLP. San Luis Potosi, Mexico.}
\email{hernan@fc.uaslp.mx}
\author{Gelasio Salazar}
\address{Instituto de F\'\i sica, UASLP. San Luis Potosi, Mexico.}
\email{gsalazar@ifisica.uaslp.mx}
\thanks{The third author was supported by CONACYT grant 106432.}

\date{\today}



\begin{document}

\maketitle

\linenumbers

\begin{abstract}
A {\em convex hole} (or {\em empty convex polygon)} of a point set $P$
in the plane is a convex polygon with vertices in $P$, containing no
points of $P$ in its interior. Let $R$ be a bounded
convex region in the
plane.  We show that the expected number of vertices of the largest
convex hole of a set of $n$ random points chosen independently and
uniformly over $R$ is $\Theta(\ttn)$, regardless of the shape of
$R$.
\end{abstract}

\section{Introduction}\label{sec:intro}
Let $P$ be a set of points in the plane. A {\em
  convex hole} (alternatively, {\em empty convex polygon}) of $P$ 
is a convex polygon with vertices in $P$, containing no points of $P$ in its interior. 

Questions about (empty or nonempty) convex polygons in point sets are
of fundamental importance in discrete and computational geometry.  
A landmark in this area is the question
posed by Erd\H{o}s and Szekeres in 1935~\cite{es1}: ``What is the smallest
integer $f(k)$ such that any set of $f(k)$ points in the plane
contains at least one convex $k$-gon?''. 
%For further results
%and references on the original Erd\H{o}s-Szekeres question we refer the reader to
%\cite{bar1,brass2}.

A variant later proposed by Erd\H{o}s himself asks for the existence
of empty convex polygons~\cite{erd1}: ``Determine the smallest
positive integer $H(n)$, if it exists, such that any set $X$ of at
least $H(n)$ points in general position in the plane contains $n$
points which are the vertices of an empty convex polygon, i.e., a
polygon whose interior does not contain any point of $X$.'' It is easy
to show that $H(3)=3$ and $H(4)=5$. Harborth~\cite{harb1} proved that
$H(5)=10$. Much more recently, Nicol\'as~\cite{nicolas} and
independently Gerken~\cite{gerken} proved that every sufficiently
large point set contains an empty convex hexagon (see also~\cite{valtr7}). It is currently known that
$30 \le H(6) \le 463$~\cite{kos,overmars}.
A celebrated construction of Horton~\cite{horton}
shows that for each $n\ge 7$, $H(n)$ does not exist. 
For further results and
references around Erd\H{o}s-Szekeres type problems, we refer the
reader to the surveys~\cite{oswin1,morris} and to the monography~\cite{brass2}.

We are interested in the expected size of convex structures in random
point sets. This gives rise to a combination of Erd\H{o}s-Szekeres type
problems with variants of Sylvester's seminal  question~\cite{syl}:
``What is the probability that four random points chosen independently
and uniformly from a convex region form a convex quadrilateral?''

Several fundamental questions have been attacked (and
solved) in this direction; see for
instance~\cite{bar3,bar4,buc1}.  
Particularly relevant
to our work are the results of Valtr, who computed exactly the
probability that $n$ random points independently and uniformly chosen
from a parallelogram~\cite{valtr1} or a triangle~\cite{valtrtriangle}
are in convex position.

Consider a bounded convex region $R$, and randomly choose $n$ points
independently and uniformly over $R$. We are interested in estimating
the expected {\em size} (that is, number of vertices) of the largest convex
hole of such a randomly generated point set.

Some related questions are heavily dependent on the shape of $R$. For
instance, 
the expected number of vertices in the convex
hull of a random point set, which is $\Theta(\log n)$ if $R$ is the
interior of a polygon, and $\Theta({n^{1/3}})$ if $R$ is
the interior of a convex figure with a smooth boundary (such as a disk)~\cite{renyisulanke,renyisulanke2}.
In the problem under consideration, it turns
out that the order of magnitude of the expected number of vertices of the largest
convex hole is independent of the shape of $R$:

\begin{theorem}\label{thm:ferr1}
Let $R$ and $S$ be bounded convex regions in the plane. Let $R_n$
(respectively, $S_n$) be a set
of $n$ points chosen independently and uniformly at random from $R$ (respectively,
$S$). Let
$\coha{R_n}$ (respectively, $\coha{S_n})$ denote the random variable
that measures the number of vertices of the
largest convex hole in $R_n$ (respectively, $S_n$). Then
$$
\EE(\coha{R_n}) = \Theta(\EE(\coha{S_n})).
$$
Moreover, w.h.p.
$$
\coha{R_n}=\Theta(\coha{S_n}).
$$
\end{theorem}

We remark that Theorem~\ref{thm:ferr1} is in line with the following
result proved by B\'ar\'any and
F\"uredi~\cite{barfur2}: the expected number of empty simplices in a
set of $n$ points chosen uniformly and independently at random from a
convex set $A$ with non-empty interior in $\real^d$ is $\Theta(n^d)$,
regardless of the shape of $A$.

Using Theorem~\ref{thm:ferr1}, 
we have determined the expected number of vertices of a largest convex
hole up to a constant multiplicative factor:

\begin{theorem}\label{thm:hole0}
Let $R$ be a bounded convex region in the plane.  
Let $R_n$ be a set
of $n$ points chosen independently and uniformly at random from $R$, and let
$\coha{R_n}$ denote the random variable that measures the number of vertices of the
largest convex hole in $R_n$. Then
$$
{\EE(\coha{R_n}) =  \Theta\biggl(\frac{\log{n}}{\log{\log{n}}}\biggr).}
$$
Moreover, w.h.p.
$$
\coha{R_n} =  \Theta\biggl(\frac{\log{n}}{\log{\log{n}}}\biggr).
$$
\end{theorem}

For the proof of Theorem~\ref{thm:hole0}, in both the lower and upper
bounds we use powerful results of Valtr, who computed precisely the
probability that $n$ points chosen at random (from a
triangle~\cite{valtr1} or from a parallelogram~\cite{valtrtriangle})
are in convex position.  The proof of the lower bound is quite
simple: we partition a unit area square {$R$ (in view of
Theorem~\ref{thm:ferr1}, it suffices to establish
Theorem~\ref{thm:hole0} for a square)} into $n /t$ rectangles such that
each of them contains exactly $t$ points, where
$t={\tn}$. Using~\cite{valtr1}, with high probability in at least one
of the regions the points are in convex position, forming a convex
hole. The proof of the upper bound is more involved. We put an $n$ by
$n$ lattice in the unit square. The first key idea is that any
sufficiently large convex hole $H$ can be well-approximated with {\em
  lattice} quadrilaterals $Q_0, Q_1$ (that is, their vertices are
lattice points) such that $Q_0\subseteq H\subseteq Q_1$ (see
Proposition 3). The key advantage of using lattice quadrilaterals is
that there are only polynomially many choices (i.e., $O(n^8)$) for each of
$Q_0$ and $Q_1$. Since $H$ is a hole, then $Q_0$ contains no
point of $R_n$ in its interior. This helps to upper estimate the area $\area{Q_0}$ of
$Q_0$, and at the same time $\area{H}$ and $\area{Q_1}$ (see Claim
B). This upper bound on $\area{Q_1}$ gives that w.h.p.~$Q_1$ contains
at most $O(\log n)$ points of $R_n$. Conditioning that each choice of
$Q_1$ contains at most $O(\log n)$ points, using
Valtr~\cite{valtrtriangle} (dividing {the $(\le 8)$-gon $Q_1\cap R$ into at most eight
triangles)} we
prove that {w.h.p.~$Q_1$ does not contain $160\log n/(\log\log n)$ points in convex
position (Claim E), so w.h.p.~there is no hole of that
  size}. A slight complication is that  $Q_1$ may not lie entirely in 
$R$; this issue makes the proof somewhat more
technical.

We make two final remarks before we move on to the proofs. As in the
previous paragraph, for the rest of the paper we let $\area{U}$ denote
the area of a region $U$ in the plane. We also note  
that, throughout the paper, by $\log{x}$ we mean the natural logarithm
of $x$.

\section{Proof of Theorem~\ref{thm:ferr1}}\label{sec:shape}

Since we only consider sets of points chosen independently and
uniformly at random from a region, for brevity we simply say that such set points
are chosen at random from this region.

\vglue 0.4 cm
\noindent{\bf Claim. }{\em For every $\alpha \ge 1$ and every
  sufficiently large $n$, }
\[\EE(\hol{R_{n}}) \ge (1/\alpha)\EE(\hol{R_{\floor{\alpha\cdot
    n}}}).
\]
\begin{proof}
Let $\alpha \ge 1$. We choose a random $\floor{\alpha\cdot n}$-point set
$R_\floor{\alpha\cdot n}$ and
a random $n$-point
set $R_n$ over $R$ as follows: first we choose $\floor{\alpha\cdot n}$ points
randomly from $R$ to obtain  $R_\floor{\alpha\cdot n}$, and then from
$R_\floor{\alpha\cdot n}$ we choose randomly $n$ points, to obtain $R_n$. 
Now if $H$ is a convex hole of $R_\floor{\alpha\cdot
    n}$ with vertex set $V(H)$, then $V(H)\cap R_n$ is the vertex set
  of
 a convex hole of $R_n$. Noting that $\EE(|V(H)\cap R_n|) =
\frac{n}{\lfloor \alpha n\rfloor} |V(H)| \ge
  (1/\alpha)|V(H)|$, the claim follows.
\end{proof}

Now the expected number of vertices of the largest convex hole in a random $n$-point set is
the same for $S$ as for any set congruent to $S$. Thus
we may assume without loss of generality that $S$
is contained in $R$. Let $\beta:=\area{R}/\area{S}$ (thus $\beta \ge
1$), and let $0 < \epsilon \ll 1$.  

Let $R_\floor{(1-\epsilon)\beta\cdot n}$ be a set of 
$\floor{(1-\epsilon)\beta\cdot n}$ points randomly chosen from $R$.
Let $m:=
|S\cap R_\floor{(1-\epsilon)\beta\cdot n}|$, {and $\alpha:=n/m$.}
{Thus the expected value of $\alpha$ is $(1-\epsilon)$, and a} standard application of Chernoff's
inequality implies that  {with probability at least
$1-e^{\Omega(-n)}$} we have
{$1 \le \alpha \le (1-2\epsilon)^{-1}$.}
Conditioning on {$m$}
means that $S_m:=
S\cap R_\floor{(1-\epsilon)\beta\cdot n}$ is a randomly chosen $m$-point set
in $S$.  

{ Since $S\subseteq R$, then every convex hole in $S_m$ is also a convex hole in
$R_\floor{(1-\epsilon)\beta\cdot n}$, and so
\begin{equation}\label{eq:ek1}
\hol{R_\floor{(1-\epsilon)\beta\cdot n}} \ge
  \hol{S_m}.
\end{equation}
From the Claim it follows that
\begin{equation}\label{eq:ek2}
\EE(\hol{R_n}) 
\ge
({(1-\epsilon)\beta)}^{-1}\EE(\hol{R_\floor{(1-\epsilon)\beta\cdot
    n}}),
\end{equation}
and that if $\alpha\ge 1$, then
$\EE(\hol{S_m}) \ge (1/\alpha)\EE(\hol{S_n})$. Therefore
\begin{equation}\label{eq:ek3}
\EE(\hol{S_m}) \ge (1-2\epsilon)\EE(\hol{S_n}),\text{\rm \hglue
  0.3 cm if \,} 1\le \alpha \le (1-2\epsilon)^{-1}.
\end{equation}
Since $1\le \alpha \le (1-2\epsilon)^{-1}$ holds with probability at
  least $1-e^{\Omega(-n)}$, 
\eqref{eq:ek1}, \eqref{eq:ek2}, and
  \eqref{eq:ek3} imply that 
$\EE(\hol{R_{n}}) \ge 
(((1-\epsilon)\beta)^{-1}
(1-2\epsilon)
\EE(\hol{S_n}) - n e^{\Omega(-n)}$. Therefore
$\EE(\hol{R_{n}}) = \Omega(
\EE(\hol{S_n}))$.} 

Reverting the roles of $R$ and $S$, we obtain 
$\EE(\hol{S_n}) = \Omega(\EE(\hol{R_n}))$, and so
$\EE(\hol{R_n}) = \Theta(\EE(\hol{S_n}))$, as claimed.

{We finally note that it is standard to modify the proof to obtain
  that
w.h.p.~$\coha{R_n} = \Theta(\coha{S_n})$.}
\hfill{$\Box$}

\section{Approximating convex sets with lattice quadrilaterals}
\label{sec:forclaimc}

For simplicity, we shall break the proof of Theorem~\ref{thm:hole0}
into several steps. There is one particular step whose proof, although
totally elementary, is somewhat long. In order to make the proof of
Theorem~\ref{thm:hole0} more readable, we devote this 
section to the proof of this auxiliary result. 

{In view of Theorem~\ref{thm:ferr1}, it will suffice to prove
Theorem~\ref{thm:hole0} for the case when $R$ is an isothetic unit area
square. In the proof of the upper bound, we subdivide $R$ into a $n$
by $n$ grid (which defines an $n+1$ by $n+1$ lattice), pick a largest convex hole $H$, and find lattice
quadrilaterals $Q_0, Q_1$ such that $Q_0\subseteq H \subseteq Q_1$,
whose areas are not too different from the area of $H$. The caveat is
that the 
circumscribed quadrilateral $Q_1$ may not completely fit into $R$; for
this reason, we need to extend this grid of area $1$ to a grid of area
$9$ (that is, to extend the $n+1$ by $n+1$ lattice to a $3n+1$ by
$3n+1$ lattice).}

We recall that a
rectangle is {\em isothetic} if each of its sides is parallel to
either the $x$- or the $y$-axis.

\begin{proposition}\label{pro:forclaimc}
Let $R$ (respectively, $S$) be the isothetic square of side length $1$
(respectively, $3$) centered at the origin. Let $n>1000$ be a 
positive integer, and let $\LL$ be the {\em lattice} $\{ (-3/2+i/3n,-3/2+j/3n) \in \real^2 \ \bigl| \ 
i,j\in\{0,1,\ldots,9n\}\}$. Let $H\subseteq R$ be a closed convex set.
Then  there exists a {\em lattice} quadrilateral (that is,
 a quadrilateral each of whose vertices is a lattice point) $Q_1$ such that
 $H\subseteq Q_1$ and $\area{Q_1} \le 2\,\area{H}+40/n$. 
Moreover, if  $\area{H} \ge \ctwo/n$, then there also exists a lattice quadrilateral
$Q_0$ such that $Q_0 \subseteq H$ and $\area{Q_0} \ge
  \area{H}/32$.
%%% Need to say that $n$ is large? Where?
\end{proposition}

We remark that some lower bound on the area of $H$ is needed in order to
guarantee the existence of a lattice quadrilateral contained in $H$, as
obviously there exist small convex sets that contain no lattice points
(let alone lattice quadrilaterals).

\begin{proof}
If $p,q$ are points in the plane, we let $\overline{pq}$ denote the
closed straight segment that joins them, and by $|\overline{pq}|$ the
length of this segment (that is, the distance between $p$ and $q$).
We recall that if $C$ is a convex set, the {\em diameter} of $C$ is $\sup\{
|\overline{xy}|   : 
x,y\in C  \}$.  {We also 
recall that a {\em supporting line} of  $C$ is a
line that intersects the boundary of $C$ and such that 
all points of $C$ are in the same closed half-plane of the line.}

\vglue 0.3 cm
\noindent {\em Existence of $Q_1$}
\vglue 0.3 cm
Let $a,b$ be a diametral pair of
$H$, that is, points such that $|\overline{ab}|$  equals the diameter of
$H$ (a diametral pair exists because $H$ is closed). Now let $\ell, \ell'$ be the supporting lines of $H$ parallel to
$\oab$.

Let $\ell_a, \ell_b$ be the lines perpendicular to $\oab$ that go
through $a$ and $b$, respectively. Since $a,b$ is a diametral pair, it
follows that $a$ (respectively, $b$)
is the only point of $H$ that lies on $\ell_a$ (respectively,
$\ell_b$). See Figure~\ref{fig:fig1}.

Let $c,d$ be points of $H$ that lie on $\ell$ and $\ell'$,
respectively. Let $J$ be the quadrilateral with vertices
$a,c,b,d$. By interchanging $\ell$ and $\ell'$ if necessary, we may
assume that $a,c,b,d$ occur in this clockwise cyclic order in the
boundary of $J$. 

Let $K$ denote the rectangle bounded by $\ell_a, \ell, \ell_b$, and
$\ell'$. Let $w,x,y,z$ be the vertices of $K$, labelled so that
$a,w,c,x,b,y,d,z$ occur in the boundary of $K$ in this clockwise
cyclic order. It follows that $\area{K} =
2\area{J}$. Since $\area{H} \ge \area{J}$, we obtain $\area{K} \le
2\area{H}$.  Let $T$ denote the isothetic square of length side $2$,
also centered at the origin. It is easy to check
that since $H\subseteq R$, then $K\subseteq T$.

Let $Q_x$ be the square with side length $2/n$ that has $x$ as one of
its vertices, with each side parallel to $\ell$ or to $\ell_a$, and
that only intersects $K$ at $x$. It is easy to see that these
conditions define uniquely $Q_x$. Let $x'$ be the vertex of $Q_x$
opposite to $x$. Define $Q_y, Q_z, Q_w, y', z'$, and $w'$
analogously.

\begin{figure}[h]
\begin{center}
\includegraphics[width=\textwidth]{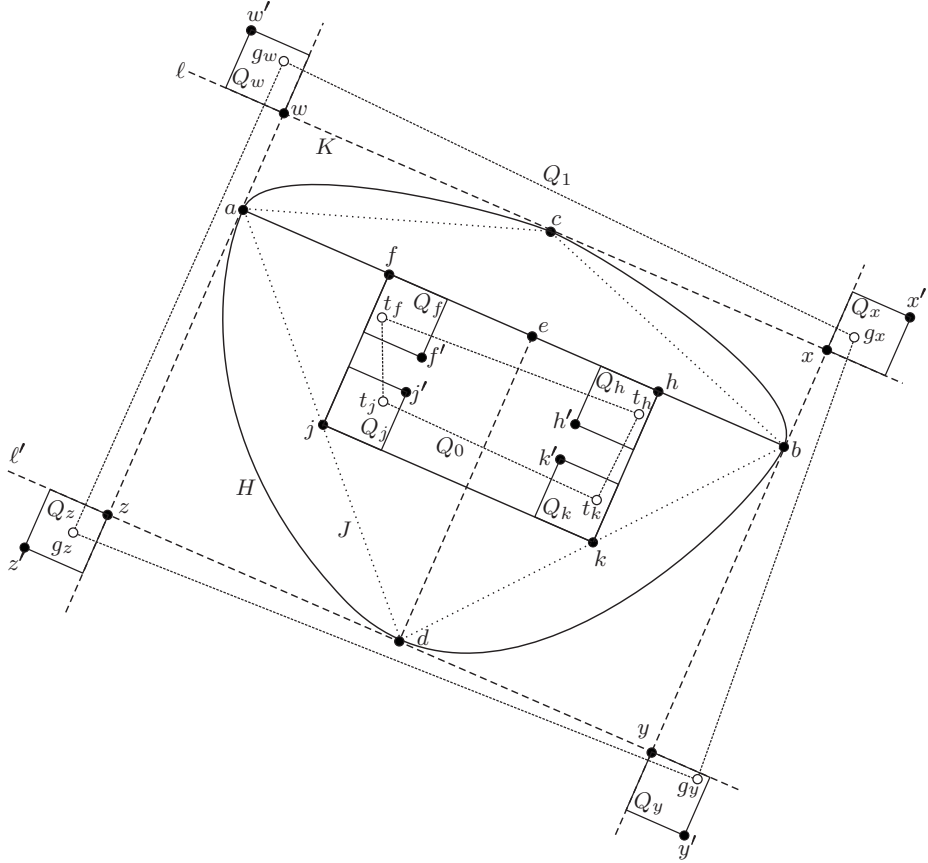}
\caption{Lattice quadrilateral $Q_1$ has vertices $g_w,$ $ g_x,$ $
  g_y,$ $g_z$, and 
lattice quadrilateral $Q_0$ has vertices $t_f,$ $ t_h,$ $ t_j,$ $ t_k$.}
\label{fig:fig1}
\end{center}
\end{figure}

Since $K\subseteq T$, it follows that $Q_x, Q_y, Q_z,$ and $Q_w$ are all contained in $S$. 
Using this, and the fact that there is a circle of diameter $2/n$ contained in $Q_x$, it
follows that there is a lattice point $g_x$ contained in the interior of
$Q_x$.  Similarly, there exist lattice points $g_y, g_z$, and $g_w$
contained in the interior of $Q_y, Q_z$, and $Q_w$, respectively.
Let $Q_1$ be the quadrilateral with vertices $g_x, g_y, g_z$, and $g_w$.

Let $\peri{K}$ denote the perimeter of $K$. The area of the rectangle
$K'$ with vertices $w',x',y',z'$ (see Figure~\ref{fig:fig1}) is $\area{K} +\peri{K}(2/n) +
4(2/n)^2$. Since the perimeter of any rectangle contained in $S$ is at
most $12$, then $\area{K'} \le \area{K} + 24/n + 16/n^2 \le \area{K} +
40/n$. Since $\area{Q_1}\le \area{K'}$, we obtain $\area{Q_1} \le
\area{K} + 40/n \le 2\,\area{H}+40/n$.

\vglue 0.3 cm
\noindent {\em Existence of $Q_0$}
\vglue 0.3 cm

Suppose without any loss of generality (relabel if needed) that the
area of the triangle $\Delta:=abd$ is at least the area of the triangle
$abc$. Since $
2\area{J}=\area{K} \ge \area{H}$ and $\area{\Delta} \ge \area{J}/2$, we have
$\area{\Delta} \ge \area{H}/4$.  By hypothesis $\area{H} \ge \ctwo/n$,
and so $\area{\Delta} \ge \ctwo/(4n)$. 

Since $a,b$ is a diametral pair, it follows that the longest side of
$\Delta$ is $\oab$. 
Let $e$ be the intersection point of $\oab$ with the line
perpendicular to $\oab$ that passes through $d$. 
Thus $\area{\Delta} = |\oab||\ode|/2$.
See
Figure~\ref{fig:fig1}.

There exists a rectangle $U$, with base contained in 
$\oab$, whose other side has length  $|\overline{de}|/2$, and such that $\area{U} =
\area{\Delta}/2$. Let $f,h,j,k$ denote the vertices
of this rectangle, labelled so that $f$ and $h$ lie on $\oab$ (with
$f$ closer to $a$ than $h$), $j$ lies on $\oad$, and $k$ lies on
$\obd$. Thus $|\ofj|=|\ode|/2$. 

Now $|\oab|<2$ (indeed, $|\oab|\le\sqrt{2}$, since
$a,b$ are both in $R$), and since $|\oab||\overline{de}|/2=\area{\Delta} \ge \ctwo/(4n)$ it follows
that $|\overline{de}| \ge \ctwo/(4n)$. Thus $|\ofj| \ge\ctwo/(8n)$. 

Now since $a,b$ is a diametral pair it follows that $\area{K} \le
|\oab|^2$. Using $\ctwo/n \le \area{H}\le \area{K}$, we obtain
$\ctwo/n \le |\oab|^2$. 
Note that
$|\ofh|=|\oab|/2$. Thus $\ctwo/(4n) \le |\ofh|^2$. 
Using $|\ofh|=|\oab|/2$ and $|\oab|\le\sqrt{2}$, we obtain $|\ofh| <
1$, and so 
$|\ofh| > |\ofh|^2 \ge \ctwo/(4n)$.

Now let $Q_f$ be the square with sides of length $2/n$, contained in
$U$, with sides parallel to the sides of $U$, and that has $f$ as one
of its vertices. Let $f'$ denote the vertex of $Q_f$ that is opposite
to $f$. Define similarly $Q_h, Q_j, Q_k, h', j'$, and $k'$.

Since $|\ofj|$ and $|\overline{fh}|$ are both at least $\ctwo/(8n)$,
it follows that the squares $Q_f, Q_h, Q_j, Q_k$ are pairwise
disjoint. Since the sides of these squares are all $2/n$, it follows
that each of these squares contains at least one lattice point. Let $t_f$
denote a lattice point contained in $Q_f$; define $t_h, t_j$, and $t_k$
analogously. 

Let $Q_0$ denote the quadrilateral with vertices $t_f, t_h, t_j$, and
$t_k$. Let $W$ denote the rectangle with vertices $f', h', j'$, and
$k'$. 

Since $|\ofj|$ and $|\ofh|$ are both at least $\ctwo/(8n)$, and the
side lenghts of the squares $Q$ are $2/n$, 
it
follows easily that $|\overline{f'h'}| > (1/2)|\ofh|$ and
$|\overline{j'k'}| > (1/2) |\ojk|$. Thus $\area{W} > \area{U}/4$. 
Now clearly $\area{Q_0} \ge \area{W}$. Recalling that $\area{U} =
\area{\Delta}/2$, $\area{\Delta} \ge \area{J}/2$, and $\area{J} =
\area{K}/2 \ge \area{H}/2$,  we obtain
$\area{Q_0} \ge \area{U}/4 \ge \area{\Delta}/8 \ge \area{J}/16 \ge
\area{H}/32$. 
\end{proof}

\section{Proof of Theorem~\ref{thm:hole0}}\label{sec:islands}

As in the proof of Theorem~\ref{thm:ferr1}, for brevity,
since we only consider sets of points chosen independently and
uniformly at random from a region, we simply say that such set points
are chosen at random from the region. 

We prove the lower and upper bounds separately.

\begin{proof}[Proof of the lower bounds]
In view of Theorem~\ref{thm:ferr1}, we may assume without any loss of
generality that $R$ is a square.  Let $R_n$ be a set of $n$ points
chosen at random from $R$. We will prove that
w.h.p.~$R_n$ has a convex hole of size at least $t$, where
$t:={\tn}$.
Let $k:={n/t}$. For simplicity, 
suppose that both $t$ and $k$ are
integers.  Let $\{\ell_0,\ell_1,\ell_2,$ $\ldots,\ell_k\}$ be a set of
vertical lines disjoint from $R_n$, chosen so that for
$i=0,1,2,\ldots,k-1$, the set $R_n^i$ of points of $R_n$ contained in
the rectangle $R^i$ bounded by $R,\ell_i$, and $\ell_{i+1}$ contains exactly
$t$ points. Conditioning that $R^i$ contains exactly $t$ points we have
that these $t$ points are
chosen at random from $R^i$. 

Valtr~\cite{valtr1} proved that the probability that $r$
points chosen at random in a parallelogram are in convex position is
$\left(\dfrac{\binom{2r-2}{r-1}}{r!}\right)^2$. Using the
bounds ${2s\choose s} \ge 4^s/(s+1)$ and   $s! \le e s^{s+1/2}
e^{-s}$, 
we obtain that this is at least
$r^{-2r}$ for all $r\ge 3$:
%\comm{I'm using Wikipedia Stirling's
% lower bound for ${2n\choose n}$ and upper bound for $n!$}
$$
\biggl(\frac{\binom{2r-2}{r-1}}{r!}\biggr)^2 \ge
\biggl(\frac{\frac{4^{r-1}}{(r-1)+1}}{er^r\sqrt{r}e^{-r}}\biggr)^2 =
\frac{(4e)^{2r}}{16e^2r^3}\cdot r^{-2r}.
$$

Since each $R^i$ is a rectangle containing $t$ points chosen at random, 
it follows that for each fixed
$i\in\{0,1,\ldots,k-1\}$, the points of $R^i_n$ are in convex position with probability at
least $t^{-2t}$. Since there are $k= n/t$ sets $R^i_n$, it follows that
none of the sets $R^i_n$ is in convex position with
probability at most 
$$(1-t^{-2t})^{n/t} \le e^{-{\frac{n}{t}}{t^{-2t}}}
= e^{-nt^{-2t-1}}.
$$

{
If the $t={\log{n}}/{(2\log{\log{n}})}$ points of an $R^i_n$ are in
convex position, then they form a convex hole of $R_n$. Thus, the
probability that there is a convex hole of $R_n$ of size 
at least ${\log{n}}/(2\log{\log{n}})$ is at least
$1-e^{-nt^{-2t-1}}$. 
Since $e^{-nt^{-2t-1}}$ $\to 0$ as
$n\to\infty$, it follows that w.h.p.~$\coha{R_n} =
\Omega({\log{n}}/(\log{\log{n}})$.
}

{
For the lower bound of 
$\EE(\coha{R_n})$,  we use once again that 
$\Pr\bigl(\coha{R_n} \ge
\log{n}/(2\log{\log{n}})\bigr) \ge 1-e^{-nt^{-2t-1}}$.
Since $\coha{R_n}$ is a non-negative random variable, 
it follows that
 $\EE(\coha{R_n})= \Omega(\log{n}/(\log{\log{n}}))$.
}
\end{proof}

\begin{proof}[Proof of the upper bounds]
%\vglue -0.2 cm
{We remark that throughout} the proof we always
implicitly assume that $n$ is sufficiently large.
We start by stating a straightforward consequence of Chernoff's
bound. This is  easily derived, for instance, from Theorem A.1.11
in~\cite{alonspencer}.

%%% {Commenting out 2nd item}.

\begin{lemma}\label{lem:alonspencer}
Let $X_1, \ldots, X_r$ be mutually independent random variables with
$\Pr(X_i=1) = p$ and 
$\Pr(X_i=0) = 1-p$, for $i=1,\ldots, r$. Let $X:=X_1 + \ldots +
X_r$. {Then, for any $s\ge r$ and $q\ge p$,}
%\begin{enumerate}
%\item 
$$
\hbox{\hglue 4.5 cm}
\Pr\bigl(X \ge (3/2) qs\bigr) < e^{-{qs/16}}.\hbox{\hglue 3.3 cm} \Box$$
%\item $P(X < \EE(X) - a) < e^{-a^2/{2\EE(X)}}$, for any $a >0$.
%\end{enumerate}
\end{lemma}

\vglue 0.3 cm

In view of Theorem~\ref{thm:ferr1}, we may assume without any loss of
generality that $R$ is a (any) square. Aiming to invoke directly
Proposition~\ref{pro:forclaimc}, we take as $R$ 
the isothetic unit area square centered at the origin, and let 
$S$ be the isothetic square of area $9$, also centered at the origin.

Let $n$ be a (large)  positive integer. 
Let $R_n$ be a set of $n$ points chosen at random from $R$. 

To establish the upper bound, we will show that w.h.p.~the largest convex
hole in $R_n$ has less than $160\,\ttn$ vertices.

Recall that $\LL$
is  the {\em lattice} $\{ (-3/2+i/3n,-3/2+j/3n) \in \real^2 \ \bigl|
\ i,j\in\{0,1,\ldots,9n\}\}$.  A point in $\LL$ is a {\em lattice
  point}.
A {\em lattice quadrilateral} is a quadrilateral
each of whose vertices is a lattice point. Now
there are $(9n+1)^2$ lattice points, and so there
are fewer than $(9n)^8$ lattice quadrilaterals in total, {and fewer
than $n^8$ lattice quadrilaterals whose four vertices are in $R$.}

\vglue 0.4 cm
\noindent{\bf Claim A. } {\em 
With probability at least {$1-n^{-{10}}$}  the random point set $R_n$ has the property that every lattice quadrilateral $Q$ with $\area{Q} < 2000\log{n}/n$ satisfies that 
$|R_n \cap Q| \le 3000\,\log{n}$.}
\vglue 0.4 cm

\begin{proof}
{Let $Q$ be a lattice quadrilateral with $\area{Q} < 2000\log{n}/n$.}
   Let $X_Q$ 
denote the random variable that measures the number of points of $R_n$
in $Q$. We apply Lemma~\ref{lem:alonspencer} with $p=\area{Q\cap
  R}\le q=(2000\log n)/n,$ 
and $ r=s=n$ to obtain $\Pr(X_Q >  3000\,\log n) < e^{-{125} \log n}=n^{-{125}}$.
 As the number of choices for $Q$ is at most $(9n)^8$, with
 probability at least {$(1- 9n^8\cdot n^{-125})  > 1-n^{-{10}}$} no {such}
$Q$ contains {more than $3000\,\log{n}$ points of $R_n$}.
\end{proof}

A polygon is {\em empty} if its interior contains no points of $R_n$.

\vglue 0.3 cm
\noindent{\bf Claim B. } {\em 
With probability at least $1-n^{-10}$ the random point set $R_n$ has the property that there is no empty lattice quadrilateral $Q\subseteq R$ with $\area{Q} \ge 20\, {\log n}/n$.} 
\vglue 0.3 cm

\begin{proof}
 The
probability that a fixed lattice quadrilateral $Q\subseteq R$ 
{with $\area{Q} \ge 20\, {\log n}/n$} is empty is
$(1-\area{Q})^n< n^{-20}.$
 Since there
are fewer than ${}n^8$ lattice quadrilaterals in $R$, it follows that
the probability that at least one of the {lattice} 
quadrilaterals {with area at least} $20\, {\log n}/n$   is
empty is less than $n^8\cdot n^{-20} {<}\,  n^{-10}$.
\end{proof}
%%%Therefore, w.h.p.~every empty lattice quadrilateral is
%%%outside $\bb$, that is, has area at most $\dcfour{\log{m}}/m$.
%%%\comm{Jozsi: one more opportunity to make fun of my probab
%%%  skills: shouldn't we also prove that w.h.p.~there exist one
%%%  (actually, many, like $\log{m}$) small empty lattice quadrilateral?}
%%% and there are many empty lattice (\le 4)-gons outside $\bb$? Does it
%%% matter? I'm not sure I'm doing the right thing in Claim B,
%%% arriving to the correct conclusion.

Let $H$ be a maximum size convex hole of $R_n$.
We now transcribe the conclusion of Proposition~\ref{pro:forclaimc}
for easy reference within this proof.

\vglue 0.4 cm
\noindent{\bf Claim C. } {\em There exists a lattice quadrilateral $Q_1$
  such that $H\subseteq Q_1$ and $\area{Q_1} \le
  2\,\area{H}+40/n$. Moreover, if $\area{H} \ge
  \ctwo/n$, then there is a lattice quadrilateral 
$Q_0$ such that $Q_0 \subseteq H$ and $\area{Q_0} \ge
  \area{H}/32$.\hfill$\Box$}  \vglue
0.4 cm

\noindent{\bf Claim D. }{\em With probability at least $1-2n^{-10}$ we
  have  $\area{Q_1} {<}\, 2000\,{\log{n}}/n$ and $|R_n\cap Q_1|\le 
3000\, {\log n}$.}
\vglue 0.4 cm

\begin{proof}
By Claim A, it suffices to show that with probability at least $1-n^{-10}$ we have that $\area{Q_1} <
2000\,{\log{n}}/n$. 

Suppose first that $\area{H} < \ctwo/n$. Then $\area{Q_1} \le
2\,\area{H} + 40/n < \cthree/n$. Since $\cthree/n <
2000\,{\log{n}}/n$, in this case we are done.

Now suppose that $\area{H} \ge \ctwo/n$, so that $Q_0$ (from Claim
C) exists.  Moreover, 
$\area{Q_1} \le 2\area{H}+40/n < 3\area{H}$.
Since $Q_0\subseteq H$, and $H$ is a convex hole of $R_n$, it follows that
$Q_0$ is empty. Thus, by Claim B,  with probability at least $1-n^{-10}$ we have that $\area{Q_0} <
20\,{\log n}/n$.
Now since $\area{Q_1} < 3\,\area{H}$ and $\area{Q_0} \ge \area{H}/32$,
it follows that 
$\area{Q_1} \le 96\,\area{Q_0}$. Thus
 with probability at least $1-n^{-10}$ we have that $\area{Q_1} \le 96\,\cdot 20\,{\log n}/n< 2000\,{\log{n}}/n$.
\end{proof}

We now derive a bound from an exact result by Valtr~\cite{valtrtriangle}.

\vglue 0.3 cm
\noindent{{\bf Claim E. }{\em The probability that $r$ points chosen at
  random from a triangle are in convex position is at most $r^{-r}$,
  for all sufficiently large $r$.}}
\vglue 0.3 cm
\begin{proof}
Valtr~\cite{valtrtriangle} proved that the probability that $r$ points
chosen at random in a triangle are in convex position is 
$2^r(3r-3)!/\bigr(((r-1)!)^3(2r)!\bigl)$. Using the bounds
    $ (s/e)^s < s! \le e\ s^{s+1/2}$ $e^{-s}$,
we obtain

\begin{equation*}
\frac{2^r(3r-3)!}{((r-1)!)^3(2r)!} <
\frac{2^r (3r)!}{(r!)^3(2r)!} \le
\frac{2^r 3
 (3r)^{3r}\sqrt{3r}e^{-3r}}{r^{3r}e^{-3r}
  (2r)^{2r}  e^{-2r} } {< {\sqrt{27r} }\, 
\biggl(\frac{27e^2}{2r^2}\biggr)^r}
< r^{-r},
\end{equation*}

\noindent where the last inequality holds for all 
 sufficiently large $r$.
\end{proof}

{For each lattice quadrilateral $Q$, the polygon $Q\cap R$ has at most
eight sides, and so it can be partitioned into at most eight
triangles. For each $Q$, we choose one such decomposition into
triangles, which we call the {\em basic} triangles of $Q$. Note that
there are fewer than $8(9n)^8$ basic triangles in total.}

\vglue 0.4 cm
\noindent{\bf Claim F. } {\em 
With probability at least $1-{2}n^{-10}$ the random point set $R_n$ satisfies that no 
lattice quadrilateral $Q$ with $\area{Q} < 2000\,{\log{n}}/n$ contains  $160\log{n}/(\log{\log{n}})$ points of $R_n$ in convex position.}
\vglue 0.4 cm

\begin{proof}

{Let $\att$ denote the set of basic triangles obtained
from lattice quadrilaterals that have area at most $2000\,\log{n}/n$.
By Claim A, with probability at least $1-n^{-10}$ every 
$T\in\att$ satisfies $|R_n\cap T| \le 3000\,\log{n}$.
Thus it suffices to show that the
probability that 
that there exists a $T\in\att$ with 
$|R_n\cap T| \le 3000\,\log{n}$ and 
$20\log{n}/(\log{\log{n}})$ points of $R_n$ in convex
position is at most $n^{-10}$.
} 

{
Let $T\in\att$ be such that
$|R_n\cap T| \le 3000\,\log{n}$, and let
$i:=|R_n\cap
T|$.
Conditioning on $i$
means that the $i$ points in $R_n\cap T$ are randomly distributed in
$T$. By Claim E, the expected number of $r$-tuples of
$R_n$ in $T$ in convex position is at most $\binom{i}{r}r^{-r}\le
\binom{3000\,\log n}{r}r^{-r} {<} (9000\,r^{-2}\log n)^r$. 
Since there are at most $8(9n)^8$ choices for $T$, 
it follows that the expected total number of such $r$-tuples (over all $T\in\att$)  with $r=20\log
n/\log\log n$ is at most $8(9n)^8 $
$\cdot  (9000r^{-2}\log n)^r<n^{-10}.$ Hence the probability that one
such $r$-tuple exists (that is, the probability that there exists a $T\in\att$ with
$20\log{n}/(\log{\log{n}})$ points of $R_n$ in convex position)
is at most $n^{-10}$.} 
\end{proof}

Now we are prepared to complete the proof of the upper bound.
 {Recall that $H$ is a maximum size
  convex hole of $R_n$, and that $H\subseteq Q_1$. It follows
immediately from Claims D and F that with probability at least
$1-4n^{-10}$ the quadrilateral $Q_1$ does not contain a set of
$160\log{n}/(\log{\log{n}})$ points of $R_n$ in convex position. 
In particular, with
probability at least $1-4n^{-10}$ the size of $H$ is at most
$160\log{n}/(\log{\log{n}})$. Therefore w.h.p.~$\coha{R_n} =
O\bigl(\log{n}/(\log{\log{n}})\bigr)$.}

{Finally, for the upper bound of 
  $\EE(\coha{R_n})$,  we use once again that with probability at least
$1-4n^{-10}$, $\coha{R_n} \le 160\,\log{n}/(\log{\log{n}})$. Since
obviously the size of the largest convex hole of $R_n$ is at most $n$, it
follows at once that 
 $\EE(\coha{R_n})= O\bigl(\log{n}/(\log{\log{n}})\bigr)$.}
\end{proof}

\section{Concluding remarks}

{The lower and upper bounds we found in the proof of
Theorem~\ref{thm:hole0} for the case when $R$ is a square (we proved
that 
w.h.p.~$(1/2)\log{n}/(\log{\log{n}}) \le \coha{R_n} \le 160\log{n}/(\log{\log{n}})$) are not
outrageously far from each other. We made no effort to optimize the $160$
factor, and with some additional work this
could be improved. Our belief is that the correct constant is
closer to $1/2$ than to $160$, and we would not be surprised if 
$1/2$ were proved to be the correct constant.}

{ There is great interest not only in the existence, but also on the
  number of convex holes of a given size (see for
  instance~\cite{baranyvaltr}). Along these lines, let us observe that
  a slight modification of our proof of Theorem~\ref{thm:hole0} yields
  the following statement. The details of the proof are omitted.  }

\begin{proposition}\label{pro:bonus}
{
Let $R_n$ be a set
of $n$ points chosen independently and uniformly at random from a square.
Then, for any positive integer $s$, the number of convex holes of $R_n$ of size
$s$ is w.h.p.~at most $n^9$.
}
\end{proposition}

{
We made no effort to improve the exponent of $n$ in this
statement.

Moreover, for ``large'' convex holes we can also give lower bounds.
Indeed, 
our calculations can be easily extended to show that for every
sufficiently small constant $c$, there is an $\epsilon(c)$ such that the 
 number of convex holes of size
at least $c\cdot\log{n}/(\log{\log{n}})$ is at most $n^8$ and at least
$n^{1-\epsilon(c)}$.
}

\section*{Acknowledgments}

This project initially evolved from conversations of the third author
with Ruy Fabila-Monroy. We thank him for
these fruitful exchanges, and for several insightful remarks and
corrections on this
manuscript. We also thank Ferran Hurtado for sharing with us his
expertise and providing guidance at the beginning of
the project. He pointed us in the right
direction by bringing to our attention the importance of settling the
(ir)relevance of the shape of the convex sets under consideration.


\begin{thebibliography}{99}

%\bibitem{os1} O. Aichholzer, F. Aurenhammer, E.D. Demaine, F. Hurtado,
%  P. Ramos, and J. Urrutia. On $k$-convex polygons. {\em Computational
%    Geometry: Theory and Applications}, {\bf 45} (2012), 73--87.

\bibitem{oswin1} O.~Aichholzer. [Empty] [colored] $k$-gons. Recent
  results on some Erd\H{o}s-Szekeres type problems. In {\em Proc.~XIII
    Encuentros de Geometr\'\i a Computacional}, pp.~43--52, Zaragoza, Spain, 2009. 

\bibitem{alonspencer} N.~Alon and J.~Spencer. The probabilistic
  method, 3rd.~Edition. Wiley, 2008.

\bibitem{barfur2} I.~B\'ar\'any and Z.~F\"uredi,  
Empty simplices in Euclidean space, {\em Canad. Math. Bull.} {\bf 30} (1987) 436--445. 

\bibitem{bar3} I.~B\'ar\'any and G.~Ambrus,
Longest convex chains,
{\em Random Structures and Algorithms} {\bf 35} (2009), 137--162.

\bibitem{bar4} I.~B\'ar\'any, Sylvester's question: the probability
  that $n$ points are in convex position, {\em Ann. Probab.} {\bf 27} (1999), 2020--2034. 

\bibitem{baranyvaltr} I.~B\'ar\'any and P.~Valtr, 
Planar point sets with a small number of empty convex polygons. 
{\em Studia Sci. Math. Hungar.} {\bf 41} (2004), 243--266. 

\bibitem{buc1} C.~Buchta, The exact distribution of 
the number of vertices of a random convex chain, {\em Mathematika}
{\bf 53} (2006), 247--254.

\bibitem{brass1} P.~Brass, Empty monochromatic fourgons in two-colored
  points sets. {\em Geombinatorics} {\bf XIV(1)} (2004), 5--7.

\bibitem{brass2} P.~Brass, W.~Moser, and J.~Pach, Research Problems in Discrete Geometry. Springer, 2005.

%\bibitem{chak} G.D.~Chakerian and L.~H.~Lange, Geometric extremum
%  problems. {\em Math. Mag.} {\bf 44} (1971), 57--69. 


%\bibitem{efron} B.~Efron, The convex hull of a random set of points,
%  {\em Biometrika} {\bf 52} (1965), 331--343.

\bibitem{es1} P.~Erd\H{o}s and G.~Szekeres, A combinatorial problem in
  geometry. {\em Compositio Math.} {\bf 2} (1935), 463--470.


\bibitem{erd1} P.~Erd\H{o}s, 
Some more problems on elementary geometry. 
{\em Austral. Math. Soc. Gaz.} {\bf 5} (1978), 52--54. 

\bibitem{gerken} T.~Gerken, 
Empty convex hexagons in planar point sets. {\em 
Discrete Comput. Geom.} {\bf 39} (2008), 239--272. 

\bibitem{harb1} H.~Harborth, Konvexe F\"unfecke in ebenen Punktmengen,
  {\em Elem. Math.} {\bf 33} (1978), 116--118.

\bibitem{horton} J.D.~Horton, Sets with no empty convex 7-gons, {\em
  Canad. Math. Bull.} {\bf 26} (1983), 482--484. 

\bibitem{kos} V.A.~Koshelev, The Erd\H{o}s-Szekeres problem. {\em
  Dokl. Math.} {\bf 76} (2007), 603--605.

\bibitem{morris} W.~Morris and V.~Soltan, The Erd\H{o}s-Szekeres
  problem on points in convex position --- a survey. {\em
Bull. Amer. Math. Soc.} {\bf 37} (2000), 437--458.

\bibitem{nicolas} C.~Nicol\'as, 
 The empty hexagon theorem. {\em Discrete Comput. Geom.} {\bf 38} (2007), 389--397.

\bibitem{overmars} M.~Overmars,
Finding sets of points without empty convex 6-gons.
{\em Discrete Comput. Geom.} {\bf 29} (2003), 153--158. 

\bibitem{renyisulanke} A.~R\'enyi and R.~Sulanke, 
\"{U}ber die konvexe H\"{u}lle von $n$ zuf\"{a}llig gew\"{a}hlten Punkten. 
{\em Z. Wahrscheinlichkeitstheorie und Verw. Gebiete} {\bf 2} (1963), 75--84.

\bibitem{renyisulanke2} A.~R\'enyi and R.~Sulanke, 
\"{U}ber die konvexe H\"{u}lle von $n$ zuf\"{a}llig gew\"{a}hlten Punkten. 
II. {\em Z. Wahrscheinlichkeitstheorie und Verw. Gebiete} {\bf 3} (1964),
138--147.

\bibitem{syl}
J.J.~Sylvester, Question 1491. {\em The Educational Times}, (London). April 1864. 

\bibitem{valtr1} P.~Valtr, Probability that $n$ Random Points are in
  Convex Position. {\em Discrete  and Computational Geometry} {\bf 13}
 (1995), 637--643.

\bibitem{valtrtriangle} P.~Valtr, The Probability that $n$ Random
  Points in a Triangle Are in Convex Position. {\em Combinatorica}
  {\bf 16} (1996), 567--573.

%\bibitem{valtr3} P.~Valtr, A sufficient condition for the existence of
%  large empty convex polygons. {\em Discrete Comput. Geom.} 
%  {\bf 28} (2002), 671--682. 


%\bibitem{valtr5} P.~Valtr, G.~Lippner, and G.~K\'arolyi,
%Empty convex polygons in almost convex sets.
%{\em Period. Math. Hungar.} {\bf 55} (2007), 121--127. 


\bibitem{valtr7} P.~Valtr, On Empty Hexagons. In: J. E. Goodman,
  J. Pach, and R. Pollack, {\em  Surveys on Discrete and Computational
    Geometry, Twenty Years Later}. Contemp. Math. {\bf 453}, AMS, 2008, pp. 433--441.



\end{thebibliography}
\end{document}